%% file: main.tex
\documentclass[sigconf, authorversion, nonacm]{acmart}

\AtBeginDocument{%
  \providecommand\BibTeX{{%
    \normalfont B\kern-0.5em{\scshape i\kern-0.25em b}\kern-0.8em\TeX}}}

\usepackage{caption}
\usepackage{subcaption}

\setcopyright{acmlicensed}
\copyrightyear{2018}
\acmYear{2018}
\acmDOI{XXXXXXX.XXXXXXX}

\acmConference[Conference acronym 'XX]{Make sure to enter the correct
  conference title from your rights confirmation emai}{June 03--05,
  2018}{Woodstock, NY}
\acmBooktitle{Woodstock '18: ACM Symposium on Neural Gaze Detection,
 June 03--05, 2018, Woodstock, NY} 
\acmISBN{978-1-4503-XXXX-X/18/06}

\begin{document}

\title{Exploring Topic Modelling of User Reviews as a Monitoring Mechanism for Emergent Issues Within Social VR Communities}


\author{Angelo Singh}
\affiliation{
  \institution{University of Glasgow}
  \country{Glasgow, Scotland}
  }
\email{2470584s@student.gla.ac.uk}

\author{Joseph O'Hagan}
\authornote{Corresponding author}
\affiliation{
  \institution{University of Glasgow}
  \country{Glasgow, Scotland}
  }
\email{joseph.ohagan@glasgow.ac.uk}

\renewcommand{\shortauthors}{Singh and O'Hagan}

\newcommand{\recroom}{\textit{Rec Room }}

\begin{abstract}
\input{tex_files/abstract}
\end{abstract}






\maketitle


\input{tex_files/introduction}

\input{tex_files/dataset}

\input{tex_files/sentiment_analysis}

\input{tex_files/word_frequency}

\input{tex_files/berttopic}

\input{tex_files/discussion}

\input{tex_files/related_work}

\input{tex_files/conclusion}



\bibliographystyle{ACM-Reference-Format}
\bibliography{references}


\end{document}

%% file: tex_files/abstract.tex
Users of social virtual reality (VR) platforms often use user reviews to document incidents of witnessed and/or experienced user harassment. 
However, at present, research has yet to be explore utilising this data as a monitoring mechanism to identify emergent issues within social VR communities. 
Such a system would be of much benefit to developers and researchers as it would enable the automatic identification of emergent issues as they occur, provide a means of longitudinally analysing harassment, and reduce the reliance on alternative, high cost, monitoring methodologies, e.g. observation or interview studies. 
To contribute towards the development of such a system, we collected approximately 40,000 \textit{Rec Room} user reviews from the \textit{Steam} storefront. 
We then analysed our dataset's sentiment, word/term frequencies, and conducted a topic modelling analysis of the negative reviews detected in our dataset. 
We report our approach was capable of longitudinally monitoring changes in review sentiment and identifying high level themes related to types of harassment known to occur in social VR platforms.

%% file: tex_files/introduction.tex
\section{Introduction}
Users of social virtual reality (VR) applications utilise user reviews as a reporting mechanism for experienced and/or witnessed harassment within social VR communities \cite{mum-user-reviews}.
Such records of user harassment range from high level insights, e.g. \textit{``some of the people that use it are sick!''} \cite{mum-user-reviews}, to more descriptive accounts of how the harassment occurred, e.g. \textit{``I've heard the N word in literally every server''} \cite{mum-user-reviews}. 
The insights contained in these reviews offer researchers and developers of social VR communities a large, continuously updated and growing, dataset of firsthand accounts of user harassment in a social VR community \cite{avatar-user-reviews}. 
Analysis of such a dataset could offer valuable insights for researchers and developers such as providing: a method of continuously monitoring \cite{mum-user-reviews} and identifying emergent issues within social VR communities \cite{avatar-user-reviews}, or a method of longitudinally analysing how harassment in a community changes, e.g. in response to a new moderation system being introduced, as new harassment trends emerge, etc. 
Therefore, there is much potential for tools which provide researchers and developers of social VR platforms with automatic monitoring mechanisms for emergent issues within a platform's community. 

Development of tools capable of monitoring social VR communities in this way would reduce the reliance of researchers and developers on alternative, higher cost, methodologies, e.g. surveys \cite{ieee-white-paper, carter-white-paper}, focus groups \cite{fiani2024pikachu}, interviews \cite{teenagers-interview, blackwell-harassment-long-paper}, in-situ observations \cite{maloney2020talking}, or combined approaches \cite{maloney-overview, flo-alt-methods}.
It could also provide a continuous information about these platforms that could be used to direct higher cost analysis methods \cite{mum-user-reviews}.
For example, a detected rise in a particular type of harassment by such a tool could motivate an in-depth exploration through observation or interviews of the specific type of harassment detected. 
Yet, at present, work exploring the development of such tools is limited. 

O'Hagan et al first proposed the development such a tool alongside work in which they manually thematically analysed 1000 \textit{Rec Room} \cite{rec-room} user reviews to investigate if/how user reviews were used to report hararssment in social VR communities \cite{mum-user-reviews}. 
Building on this work, Dong et al explored accounts of avatar documented in user reviews by conducting a targeted, topic modelling \cite{nlp-methods} analysis of 6000 \textit{VRChat} \cite{vrchat} user reviews \cite{avatar-user-reviews}. 
Others, meanwhile, have utilised user reviews and a natural language processing techniques \cite{nlp-methods} to investigate usability \cite{user-review-detection} or accessibility \cite{accessibility-reviews} issues in applications, to analyse emotional content \cite{bertopic-emotions}, or understand changing player behaviours \cite{topic-modelling-pokemon-go}.

In this paper we contribute an exploration of natural language processing techniques to conduct large-scale analysis of user reviews to monitor user harassment in social VR communities. 
We captured a dataset of approximately 40,000 \textit{Rec Room} user reviews from the \textit{Steam} \cite{steam} storefront dating from June 2016 up to and including February 2024.
We conducted a sentiment analysis of these reviews, observing sentiment in \textit{Rec Room} to be decreasing over time and labelling the reviews as being of positive, negative, or uncertain sentiment. 
We then conducted a word/term frequency analysis, identifying frequent references to children, verbal harassment, and racism in our negative reviews.
Finally, we conducted a BERTopic modelling of our reviews to cluster reviews detected to have similarities in their text content together, to group our reviews and identify emergent themes.  
This identified eight high level themes within our negative reviews, demonstrating our approach is effective for reducing a large data of reviews into clusters of similar topics but also motivating future work to explore how more detailed insights into the contents of these clusters might automatically be generated. 

%% file: tex_files/dataset.tex
\section{Our Data Sources}

\subsection{Why Capture \textit{Rec Room} Reviews?} 
\textit{Rec Room} \cite{rec-room} was chosen as it is a popular social VR platform with more than 75 million downloads \cite{rec-room-millions}.
It is a social VR platform that is often used when investigating player experience and harassment in social VR (e.g. \cite{blackwell-harassment-short-paper, women-social-vr}). 
It was also chosen by O'Hagan et al in their investigation into the documenting of  witnessed and/or experienced user harassment in user reviews \cite{mum-user-reviews}, and so ensured that \textit{Rec Room} user reviews were a viable data source for our study.

\subsection{Why Capture \textit{Steam} User Reviews?} 
We captured \recroom user reviews from June 2016 up to and including February 2024 from the \textit{Steam} online storefront \cite{steam}. 
We chose \textit{Steam} as it is the predominant PC gaming storefront, and so the largest available single source of \recroom user reviews. 
It also provides developers/researchers access to the \textit{Steamworks API} which can be used to conduct analysis on a \textit{Steam} application's user reviews. 
Additionally, \textit{Steam} requires users be owners/players of an application before they can leave a review on it \cite{steamFAQ}, reducing rates of fake reviews within any captured dataset. 
It should also be noted that \recroom, although marketed as a social VR application and analysed as a social VR by researchers \cite{maloney-overview, blackwell-harassment-long-paper}, enables cross-platform play between non-VR (PC, console, and smartphone) and VR users \cite{rec-room-steam}. 
Where other storefronts are limited to reviews of specific platform users, e.g. Quest users on Meta's storefront \cite{rec-room-store}, the \textit{Steam} storefront enables the capture of reviews from both VR and PC users, a more representative set of reviews from \textit{Rec Room's} user base. 
The insights of this a mixed dataset (containing reviews from VR and PC users) are of equal relevance to any potential VR user, however, as any described interaction between users could potentially be experienced and/or witnessed by a VR user, given the cross-platform nature of application.

\section{Data Capture \& Pre-processing}
We created a scraper using Python and the \textit{Steamworks API} to create a dataset containing all English language user reviews on \textit{Rec Room's Steam} store page. 
Reviews were collected from most recent to oldest. 
We ran the scraper in February 2024, creating a dataset of approximately 40,000 user reviews ranging from June 2016 (\textit{Rec Room's} release month) up to, and including, February 2024.

In topic modelling and clustering tasks, inaccurate results can emerge when analysed texts do not provide sufficient or meaningful detail for accurate classification \cite{cleaningforml}. 
While the \textit{Steam} storefront requires users be owners/players of an application prior to reviewing it \cite{steamFAQ}, the review content itself is not standardised. 
As such, reviews can vary considerably in their content, e.g. being meaningful in-depth discussions of a user's experience, being only a few words such as \textit{``fun game''}, being pieces of ASCII artwork, or being spam of random characters, emojis, or repeated sequences of words, etc. 
Therefore, the decision was made to filter the dataset to exclude reviews likely to add noise and result in inaccurate results emerging.
Additionally, factors such as spelling mistakes, case inconsistencies, and special characters can also introduce inaccuracies in the results of classification models \cite{mishra2020dqi}. 
In addition to filtering, the dataset was also cleaned to enforce a standardisation of review format.

\subsection{Data Filtering}
A filtering of the data was conducted to remove reviews based on the \textit{presence of spam reviews}, and to group the remaining reviews by their \textit{length}. 
To address the presence of spam reviews, e.g. reviews containing ASCII art or the repetition of a single word, phrase or character repeated for the entire review, we created a script to search and exclude such reviews.
This script searched for reviews containing repeated consecutive special characters (set to 6 consecutive characters) to find ASCII artworks. 
It also searched for spam character/word(s) reviews by investigating the ratio of unique characters/words to their frequency within the review text, e.g. a review containing a small number of unique characters or words but a high frequency rate would be flagged as a potential spam review.  

The reviews were then sorted into 3 groups based on their length. 
Inspecting the length of the remaining reviews approximately: 50\% of the reviews were composed of \textit{``5 or less words''}, 44\% \textit{``between 6 and 50 words''}, and 6\% \textit{``more than 50 words''}. 
These groups were chosen as texts greater than 50 words have been shown to not benefit clustering algorithms as more semantic meaning can be captured by the additional words, making the grouping process more ambiguous. Similarly, too few words, e.g. 5 or less, can introduce similar ambiguity into results from too little semantic meaning to work from \cite{cleaningforml, mishra2020dqi}. 
The group of reviews of length \textit{``between 6 and 50 words''} (approximately 17,000 user reviews) was then extracted out as the analysed set of reviews in our work.

\subsection{Data Cleaning \& Pre-Processing}
The reviews for analysis were then cleaned in order to obtain a standardised format of review which could be further analysed. 
The widely used Python library \textit{Natural Language Toolkit (NLKT)} was used to perform the data cleaning. 
The data was cleaned to remove stop-words, e.g. frequently used words that carry no useful meaning such as ``the'', ``and'', ``to''. Removing such words ensures analysis using supervised or unsupervised approaches can assign the right importance to the meaningful words and reduces noise added by incorrectly adding importance to unnecessary, non-meaningful words which occur often \cite{stopwords}. 
A tokenisation was performed to break the review texts into tokens representing meaningful sequences of characters (in our case corresponding to word(s)) \cite{stopwords}. 
Finally, a stemming was performed to normalised the review texts by reducing the words to their base root forms \cite{stemming}. 
This process is necessary/beneficial for our analysis of the texts, e.g. to aid clustering techniques to aid different forms of a word to be grouped.
For example, the word \textit{``Quick''} can be extracted from words such as \textit{``quickly''} and \textit{``quicker''} facilitating clustering and topic modelling techniques to count multiple representations of a word as one \cite{stemming}. 

%% file: tex_files/sentiment_analysis.tex
\section{Sentiment Analysis}
Despite \textit{Steam} providing an overall summary recommendation of \textit{Rec Room's} user reviews (e.g. \textit{``Overall reviews: Very Positive''}) it does not show how review sentiment has changed over time. 
Being able to examine this, however, would provide insight into how sentiment of \recroom has changed since its initial release, e.g. as the developers introduce new features and as the player base has grown over time.
Additionally, by analysing our reviews sentiment, we would obtain a subset of identified \textit{``negative reviews''} which could be used to conduct more targeted analysis of. 

Although \textit{Steam} user reviews are composed of a binary \textit{``Recommended''} or \textit{``Not Recommended''} rating alongside the review text, prior work has shown that such shorthand ratings do not accurately reflect users' review text content/sentiment \cite{mum-user-reviews}.
For example, the review text or rating might be used sarcastically \cite{goodSarcasm}. 
As such, it is necessary that sentiment of the user reviews be determined based on the review text alone.

\subsection{Manual Labelling a Subset of User Reviews}
We used a supervised learning approach to assign a sentiment to each review.
This decision was motivated by the absence of labels for the captured dataset and its size, although 100 user reviews were manually labelled by the researchers in order to assess the performance of the pre-trained model used (Section \ref{subsection:model-performance-assessment}). 
A single researcher performed the labelling, performing two cycles whilst labelling the data. 
Another researcher then reviewed the labels with any differences in labels being discussed and resolved by the researchers. 

The reviews were labelled/classified into three classes: \textit{Positive, Negative,} and \textit{Neutral} sentiment - a multi-class sentiment analysis. 
Although a binary classification (labelling as being \textit{Positive} and \textit{Negative} only) may increase the accuracy of a sentiment analysis model generally \cite{numberoflabelsclassification}, the nature of user review text is often mixed in sentiment outlining both positive and negative aspects \cite{mum-user-reviews}. 
Therefore, the decision was made to include a \textit{Neutral} classification to handle instances where review sentiment was mixed as such an additional classification can increase confidence in the sentiment analysis by establishing a clearer distinction between positive, negative, and mixed sentiment reviews. 

\textit{\textbf{Limitations}:} We acknowledge our supervised learning approach is subject to some inaccuracy given the generalised approach in which the model used is trained. 
Meaning, it has not been trained and cannot be used to search specifically for issues such as \textit{``harassment in social VR''} in the dataset. 
However, it is sufficient for the purpose of conducting a sentiment analysis of our data. 
Furthermore, to show the accuracy and precision of our analysis we conducted an evaluation of the model used in our approach using a manually labelled subset of our data.

\subsection{Identifying a Pre-Trained Model for Sentiment Analysis}
\label{subsection:model-performance-assessment}

Two widely used pre-trained models (\textit{cardiffnlp} and \textit{lxyuan}) were selected from the \textit{Hugging Face} library \cite{huggingfacemodels} as potential models to conduct our sentiment analysis:  
\begin{itemize}
    \item \textit{cardiffnlp/twitter-roberta-base-sentiment-latest} \cite{huggingface-xlm-roberta} 
    \item \textit{lxyuan/distilbert-base-multilingual-cased-sentiments-student} \cite{huggingface-lxyuan}
\end{itemize}
To determine which model to use, we evaluated the performance of against a our dataset of 100 labelled user reviews. 
The confusion matrices of the \textit{cardiffnlp} and \textit{lxyuan} models indicated better performance for the \textit{cardiffnlp} model. 
The performance of both models was then evaluated in terms of their: \textit{accuracy}, \textit{precision}, \textit{recall}, and \textit{F1-score}.

Table \ref{table:model-performance-review} summarises this performance review for both models. 
Overall, \textit{cardiffnlp} was found to perform better than \textit{lxyuan}. 
The \textit{cardiffnlp} was found to have a higher accuracy score, indicating it managed to correctly identify more labels, in addition to higher values of precision, recall, and F1-score. 
Therefore, the decision was made to perform the sentiment analysis of the user reviews using the \textit{cardiffnlp} for training.

\begin{table}[b!]
\centering
\begin{tabular}{llccc}
\hline
\textbf{Metric} & \textbf{True Labels} & \textbf{\textit{cardiffnlp}} & \textbf{\textit{lxyuan}} \\
\hline
\textbf{Accuracy:} & Overall & 0.73 & 0.61 \\ \\
\hline
\textbf{Precision:} & Positive & 0.78 & 0.71 \\
                           & Neutral & 0.42 & 0.00 \\
                           & Negative & 0.70 & 0.45 \\
\hline
\textbf{Recall:} & Positive & 0.81 &  0.77 \\
                        & Neutral & 0.19 & 0.00 \\
                        & Negative & 0.88 & 0.63 \\
\hline
\textbf{F1 Score:} & Positive & 0.80 &  0.74 \\
                          & Neutral & 0.16 & 0.00 \\
                          & Negative & 0.78 & 0.53 \\
\hline
\end{tabular}
\caption{Our performance evaluation of both models indicate \textit{cardiffnlp} is better suited for our analysis, and could more accurately classify the sentiment of our labelled dataset.} 
\label{table:model-performance-review}
\end{table}

\subsection{Results: Review Sentiment Over Time}
\begin{figure}[b!]
\centering
    \includegraphics[width=1\linewidth]{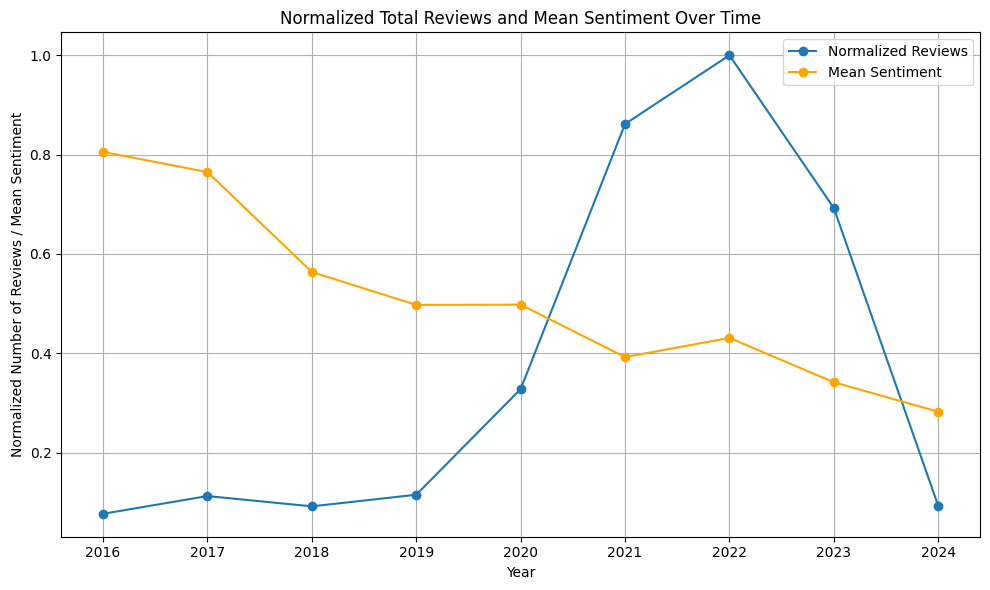}
    \caption{Comparison of mean sentiment over time (ranging +1 to -1) and normalised number of reviews from 2016 to 2024. A downward trend in sentiment is seen since \textit{Rec Room's} initial 2016 release, more than halving in its mean sentiment score score by 2023.}
    \label{fig:sentiment_over_time}
\end{figure}

Applying a numerical value of +1 (positive), 0 (neutral), and -1 (negative) to the sentiment of each review, a mean value of sentiment can be obtained from the dataset. 
Figure \ref{fig:sentiment_over_time} shows this mean value, plotted against the number of reviews, for each year in our dataset. Note: the values for 2024 represent only the first 2 months of 2024 but does indicate a continued downward trend in sentiment score.    

\begin{figure*}[]
  \centering
  \includegraphics[width=0.7\textwidth]{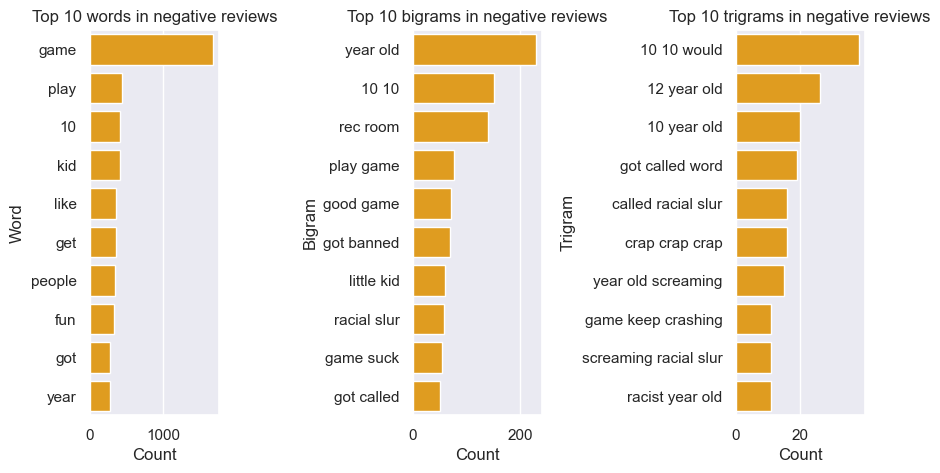} 
  \caption{The 10 most frequent of individual words, bigrams and trigrams (from left to right) in the negative reviews dataset. References to children, verbal harassment, and racism occur across all counts.}
  \label{figure:bag-of-words}
\end{figure*}

Figure \ref{fig:sentiment_over_time} shows an observable downward trend in review sentiment each year since its 2016 release. 
This trend has continued despite a significant increase in the number of user reviews posted since 2020. 
This provides some indication of how \textit{Rec Room's} users have responded to changes made to \recroom by its developers year-on-year. That is, any changes made (e.g. in response to negative community feedback, as the application continues to evolve over time) have failed to increase users' sentiment of their experience. 
While it could be argued that despite this decrease \textit{Rec Room's} overall sentiment remains positive (e.g. its \textit{Steam} store page listing it as rating \textit{``Very Positive''}), the number of reviews posted in 2023 was lower than 2021 and 2022 suggesting a potential peak in \recroom acquiring new players year-on-year. 
Furthermore, sentiment that continues to grow more negative amongst players will likely result in both existing users quitting from regular use and difficulty in attracting/retaining new users. 

Our sentiment analysis identified 3920 user reviews as being of negative sentiment.
While this represents a significant decrease from our analysed data set (of approximately 17,000 reviews), such a decrease is not unexpected due the inclusion of a neutral classification in our approach and prior works indicating high rates of reviews of mixed sentiment \cite{mum-user-reviews}. 
Additionally, such a significant decrease was also found in O'Hagan et al's analysis of 1000 user reviews of which they identified only 114 contained content relevant to their evaluation \cite{mum-user-reviews}.

%% file: tex_files/word_frequency.tex
\section{Term Frequencies and Importance}
After conducting the sentiment analysis, and obtaining a labelled subset of 3920 negative reviews, we then investigated what themes, if any, could be observed within this negative reviews dataset. 
Identifying such themes would provide insights into why users discussed their experience with \recroom as being negative in their review, e.g. describing an experience of harassment within a social VR or discussing usability issues with the application, etc. 
To explore this, we examined the frequency of words/terms within our negative reviews dataset using a bag-of-words (with n-gram representation) and their importance using TF-IDF scores.

\subsection{Bag-of-Words and N-Gram Representation}
Bag-of-words is a common approach used for text analysis to provide an overview of which unordered words occur most often in a dataset. 
However, due to its analysis of unordered words, with this approach there is no weighting/consideration given to the semantic meaning of words when combined with other words.
In the context of analysing user reviews this is a limitation. 
Therefore, to address this limitation, we used n-grams (sequences formed by n sequential terms within a document) to capture sequential relationships with words adjacent to them. 
This provided further context and meaning for our analysis than individual words alone.
We chose to analyse bi-grams (2 consecutive words) and tri-grams (3 consecutive words) in our analysis.

\subsection{Results: Bag-of-Words and N-Grams Counts}
\label{section:results-bag-of-words}
Figure \ref{figure:bag-of-words} shows the ten most frequent individual words, bigrams, and trigrams in the negative labelled reviews. 
While \textit{``game''} being the most frequent word is not surprising, it does highlight that users think of \recroom as a \textit{``game''} rather than a social VR platform, in-line with prior work highlighting how consumer mental models of VR consider the technology to be a gaming device \cite{mmve-joseph}. 
References to \textit{``play''} and \textit{``fun''} within the negative reviews may be due to reviews in which users praise the \textit{Rec Room's} design but go on to state their experience using \recroom is impacted by other users online. 
Such reviews for \recroom have been identified by O'Hagan et al who found a significant portion of \recroom user reviews argued it was \textit{``a good, well designed game that their experience with was severely impacted by a bad community and their experience with the online player base''} \cite{mum-user-reviews}. 
However, it could also be due to reviews which identify a negative point and state it would be more fun to play if this was addressed.  

References to \textit{``kid''} and \textit{``year''} (clarified in the bigrams and trigrams as \textit{``little kid''} and \textit{``N year old''}) highlight the role of younger users in \textit{Rec Room's} reviews.
The trigrams: \textit{``racist year old''} and \textit{``year old screaming''}, provide further clarity on the role they play, that is, user reviews featuring reports of younger users verbally harassing other users (an observation often reported in prior works \cite{maloney-child-1, maloney-child-2, blackwell-harassment-long-paper}). 
Further reference to verbal harassment within the negative reviews is also found in the bigrams and trigrams: \textit{``racial slur''}, \textit{``got called''}, \textit{``got called word''}, \textit{``called racial slur''}, and \textit{``screaming racial slur''} - all of which follow the model of how verbal harassment is reported to occur in \recroom in past research \cite{blackwell-harassment-long-paper, maloney-overview, mum-user-reviews}. 

Of note also is the occurrence of high level critiques, e.g. \textit{``good game''}, \textit{``game suck''} and \textit{``crap crap crap''}, which match the depth of feedback often reported as occurring in user reviews \cite{avatar-user-reviews, mum-user-reviews}. 
That is, users often will indicate a positive or negative sentiment without providing detail indicating why this is felt. 
The inclusion of \textit{``got banned''} meanwhile is unclear from this analysis alone if it is in reference to users complaining they were banned (accurately or inaccurately) or if they are indicating certain user behaviours should be banned. 
Regardless, it does indicate discourse amongst users regarding the exclusion of individuals as a moderation technique (in response to negative user behaviours). 
The inclusion of \textit{``game keep crashing''} is of note as the reporting of technical issues was reported also in Dong et al's analysis of \textit{VRChat} user reviews \cite{avatar-user-reviews}. 
Finally, it should be noted, there are some occurrences which are unclear, e.g. \textit{``10 10''} and \textit{``10 10 would''}.

\subsection{TF-IDF}
We used TF-IDF to evaluate the importance of words used within the review texts. 
While term frequency is an important factor when examining our review texts, it is important also to consider the \textit{``importance''} of words within our texts. 
For example, a word such as \textit{``racist''} may not appear in every review, and so would have a lower frequency count, but might appear often within the reviews that use it - indicating a high importance when used. 
Similarly, commonly used words with high frequency but low importance can be filtered out.

\begin{figure}[b!]
  \centering
  \includegraphics[width=0.375\textwidth]{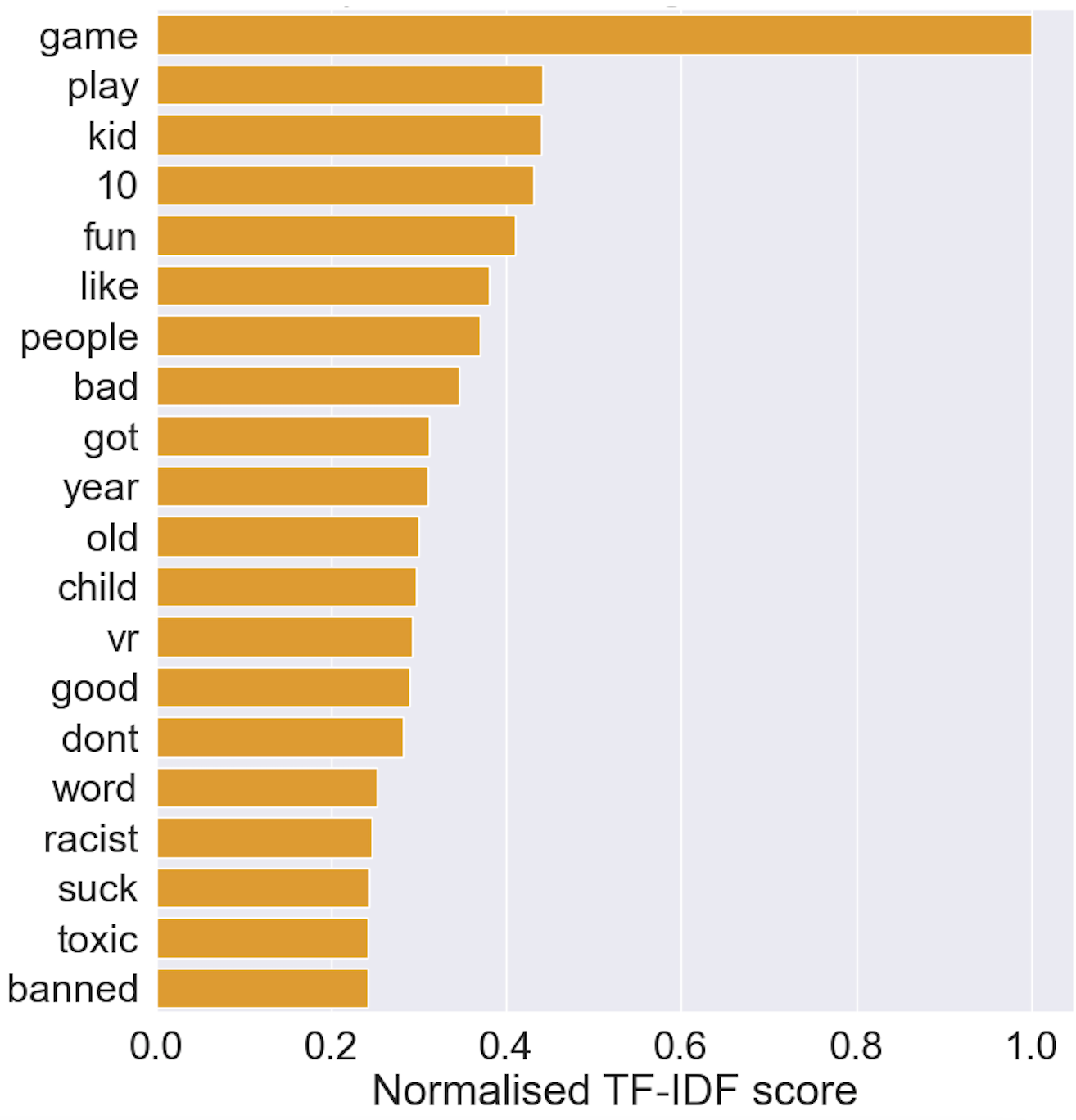} 
  \caption{The 20 highest TF-IDF scores in our negative reviews. Terms referring to children, racism, toxicity, and negative sentiment were prevalent.}
  \label{figure:tf-idf_scores}
\end{figure}

\subsection{Results: TF-IDF Scores}
Figure \ref{figure:tf-idf_scores} shows the twenty highest scoring words by their TF-IDF scores in the negative labelled reviews. 
As in the bag-of-word count of individual words (Figure \ref{figure:bag-of-words}), \textit{``game''} scored highest. 
While expected, due to its high frequency of use throughout many user review texts, as outlined in Section \ref{section:results-bag-of-words}, it is worth noting users' reference to \recroom as a game rather than a social VR platform or some other description. 

The words \textit{``play''} and \textit{``fun''} scored highly, again suggesting the potential for reviews to be structured as praising \textit{Rec Room's} design but then identify some aspect of it (e.g. user harassment) as impacting the experience \cite{mum-user-reviews}.  
The words \textit{``kid''} and \textit{``child''} highlight the prevalence of children/younger users within the context of negative reviews, a result in-line with prior works \cite{maloney-child-1, fiani2024pikachu}. Additionally the words \textit{``year''} and \textit{``old''}, combined with the insights given by the bigrams and trigrams, provides additional support for the frequent discussion of younger users within the negative reviews. 

The words \textit{``racist''} and \textit{``toxic''} highlight reports of verbal harassment within the negative reviews which O'Hagan et al identify as the primary type of harassment reported in \textit{Rec Room's} user reviews \cite{mum-user-reviews}. 
Additionally, the words \textit{``got''} and \textit{``word''}, combined with the context given by the bigrams and trigrams, provide additional support for this, e.g. phrases such as \textit{``got called racist word''} when documenting witnessed and/or experienced verbal harassment. 

Single words used as general high level critiques are also prevalence, the words \textit{``like''}, \textit{``bad''}, and \textit{``suck''}. 
The inclusion of \textit{``people''} is noteworthy, highlighting discourse surrounding other users within the context of negative reviews, e.g. other players impacting a user's experience with \recroom \cite{mum-user-reviews, maloney-overview, blackwell-harassment-short-paper}. 
While the inclusion of \textit{``banned''} again highlights discourse in reviews wherein the exclusion of individuals is discussed. 
Finally, as in the bag-of-words approach, due to the analysis method there are some words which are unclear in their usage, e.g. \textit{``10''} and \textit{``dont''}.

%% file: tex_files/berttopic.tex
\begin{figure*}[h]
\centering
\includegraphics[width=0.9\textwidth]{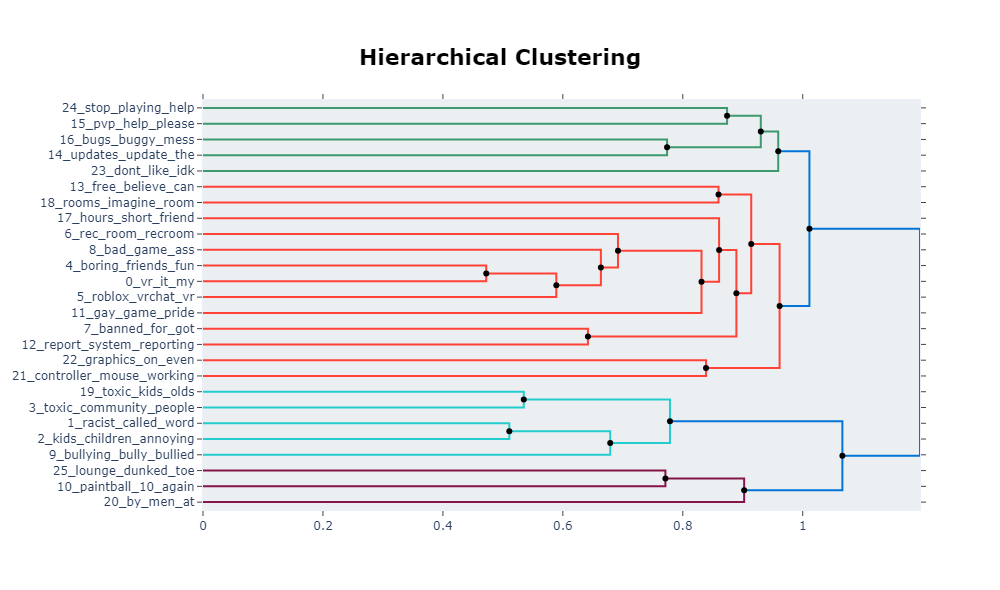} 
\vspace{-7mm}
\caption{A visualisation of the hierarchical representation of topics when clustering our dataset of negative reviews only.}
\vspace{-2mm}
\label{figure:bertopic-negative-hier}
\end{figure*}

\begin{figure}[h]
\centering
\includegraphics[width=0.5\textwidth]{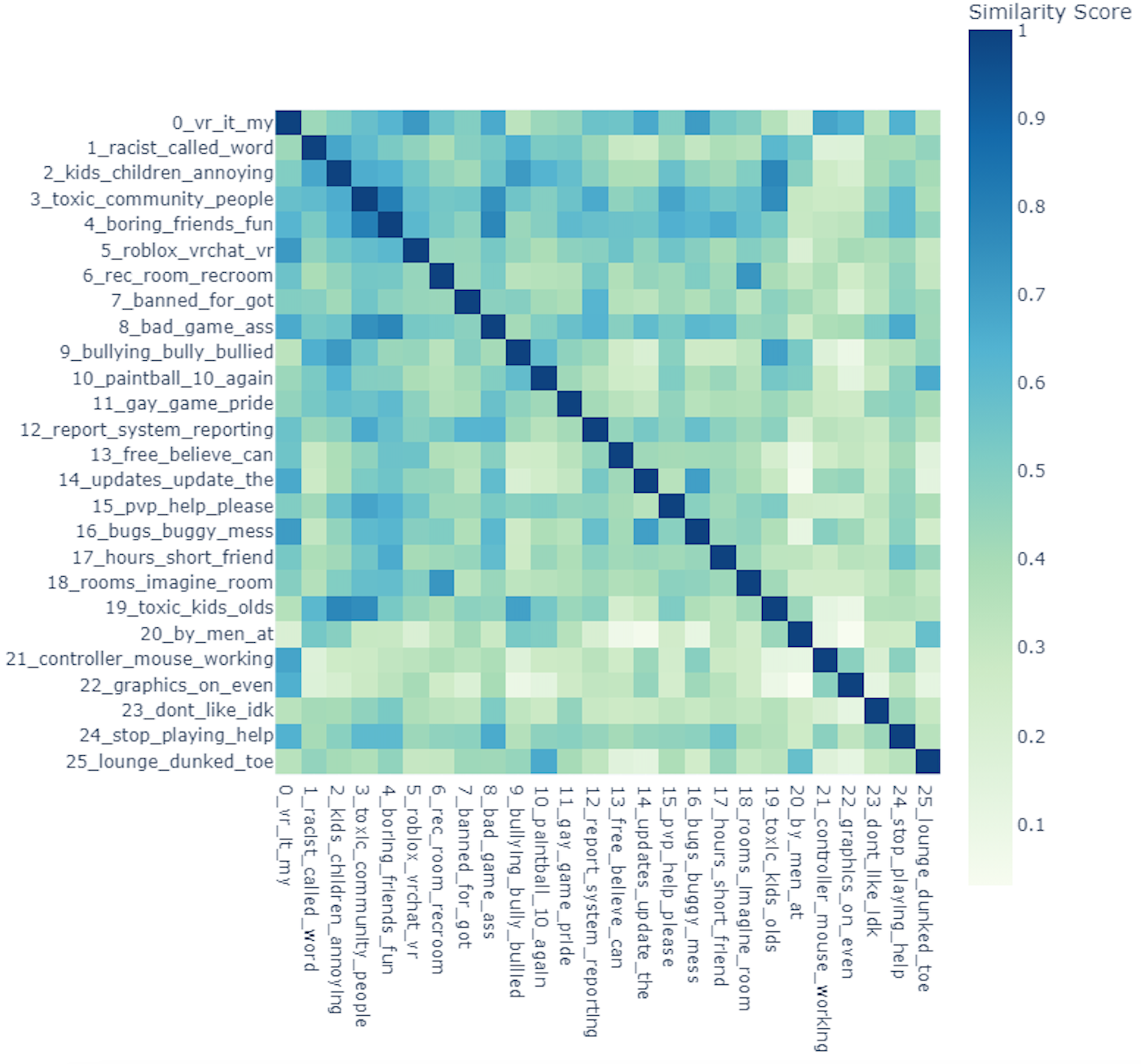} 
\vspace{-4mm}
\caption{A heatmap representation of topic similarity over identified topics. Darker patches on the heat map indicate a higher similarity between topics. This and the visualisation representation were used to identify the emergent from the BERTopic analysis.}
\vspace{4mm}
\label{figure:bertopic-negative-heatmap}
\end{figure}

\section{Identifying Emergent Themes Within User Reviews}

To investigate what themes could be observed within our dataset we employed BERTopic modeling to create dense clusters of thematically similar review texts. 
While the word/term frequency and importance analysis of our dataset provided high level insights into the words/terms used throughout the reviews, and some insight into the content/themes within the reviews, additional work is necessary to explore the review content in more detail. 

The use of BERTopic is well suited to such an analysis and is a widely used approach for this within the literature, e.g. \cite{avatar-user-reviews, rogers2021primer, bertopic-emotions}. 
Unlike other analysis techniques (e.g. Latent Dirichlet Allocation or K-Means Clustering) BERTopic does not require a pre-defined number of topics/themes to cluster by, hence it is most applicable to our approach where the number of topics is unknown. 
Moreover, BERTopic is more flexible and can be customised to best fit specific analysis tasks, offering a modular framework that allows for the use of many methodologies for embedding, reduction, and clustering.

\textit{\textbf{Specifics of our approach:}} For tokenisation \textit{CountVectorizer} of scikit-learn \cite{scikit_learn} was used, as it is a widely used approach and one that is regularly updated with out-of-vocabulary words. 
To assign weights to the terms, the weighting scheme \textit{c-TF-IDF} was used. 
For the embedding step, given the high availability of sentence transformers within BERTopic, the sentence-transformers model \textit{``all-MiniLM-L6-v2''} was used.
For dimensionality reduction, the \textit{UMAP} (Uniform Manifold Approximation and Projection) technique was used.
Finally, for the clustering of the data a preliminary analysis determined the \textit{DBSCAN} (Density-Based Spatial Clustering of Applications with Noise) technique was not an effective technique for our dataset. 
Therefore, the \textit{HDBSCAN} (Hierarchical Density-Based Spatial Clustering of Applications) technique was used to perform our analysis instead (and was found to be a more effective method for doing so).

\subsection{Our BERTopic Modelling Analysis}

We conducted a BERTopic modeling analysis of our negative review subset (3920 reviews). 
Generating the topics found within these produced 25 clusters of reviews. 
Figure \ref{figure:bertopic-negative-hier} provides a visualisation of their hierarchical topic representation and keywords summarising each topic. 
While some noise remains and some clusters are irrelevant to our investigation, e.g. \textit{``rec-room-recroom''}, or unclear, e.g. \textit{``free-believe-can''}, there are significantly more relevant clusters than in our clustering of all of the reviews. 

To further investigate these clusters of themes identified within our negative reviews, we performed a manual inspection and merging of the clusters into themes of interest for further discussion. 
In doing this, we would identify negative issues reported by many users across the negative reviews, and so identify issues that are brought up and/or discussed by many users in their negative reviews of \recroom. 
To do this, we generated a heatmap of the topic similarity over the identified topics (Figure \ref{figure:bertopic-negative-heatmap}). 
This, combined with the hierarchical topic representation was used to identify sufficiently similar clusters to merge into thematically similar themes.

\subsection{Themes Identified By Our Analysis}

We identified 8 themes from our BERTopic analysis:

\subsubsection{\textbf{Themes Concerning \textit{Rec Room's} Community:}}
We identified 2 themes in the negative reviews concerning \textit{Rec Room's} community and users. 
These were: \textit{A Toxic Community (162 reviews)}, and \textit{Problematic Child Users (205 reviews)}.
Such themes are in-line with past works which have identified \textit{Rec Room's} community as a source of much issue for other users, e.g. through users harassing other users \cite{mum-user-reviews, maloney-overview, blackwell-harassment-long-paper}.  
As reported by prior works also, e.g. \cite{maloney-child-1, maloney-child-2, mum-user-reviews}, the role of children as harassers was found in our analysis, with keywords such as \textit{``annoying children''} and \textit{``toxic children''} frequently being used within the negative reviews to describe their behaviour towards other users. 
However, less prevalent in our analysis was the identification of children as victims of harassment, despite the findings of O'Hagan et al indicating the reporting of such accounts in user reviews \cite{mum-user-reviews}.

\subsubsection{\textbf{Themes Concerning How User Harassment Occurs in \textit{Rec Room}:}}
We identified 3 themes in the negative reviews concerning how harassment of users occurs. 
These were: \textit{Racism (207 reviews)}, \textit{Bullying (66 reviews)}, and \textit{LGBTQ+ Harassment (32 reviews)}.
We again report references harassment being done verbally across these themes, e.g. clusters such as \textit{``racist-called-word''} emphasising the verbal nature of the harassment. 
This result is in-line with past work which has identified verbal harassment to be the predominant type of harassment reported in the user reviews of social VRs \cite{mum-user-reviews}. 
Additionally, these themes are said by prior works to be prominent forms of harassment in social VRs \cite{blackwell-harassment-long-paper, blackwell-harassment-short-paper, maloney-overview}, and our work affirms the continued presence, and longstanding nature of, such issues within \textit{Rec Room's} community.

\subsubsection{\textbf{Themes Concerning Player Moderation and Technical Issues:}}
We identified 3 themes in the negative reviews concerning player moderation and technical issues that occur. 
These were: \textit{Banned Users (98 reviews)}, \textit{A Faulty Reporting System (19 reviews)}, and \textit{Technical Issues (14 reviews)}.
Discussion of \textit{Banned Users} and \textit{A Faulty Reporting System} highlights the role that exclusion/banning currently plays in managing problematic users. 
However, it also highlights a limitation of this mitigation approach - that is, the potential for accidental banning of a user who does not deserve it, or not banning a user who does. 
Complaints about this in user reviews has been observed in past work which has noted the use of user reviews as a method for advocating for better user moderation and reporting systems \cite{mum-user-reviews}. 
The reporting of other \textit{Technical Issues} in user reviews is noted from our word/term frequency analysis where we found frequent use of the trigram \textit{``game keep crashing''}. 
While our analysis approach does not enable us to comment further on the nature of the technical issues experienced, instead only allowing us to identified prevalent reporting of such issues in user reviews, it is noteworthy that user reviews are used to report such issues by users. 

%% file: tex_files/discussion.tex
\section{Discussion}
We used sentiment analysis, word/term frequencies and importance, and BERTopic modelling to analyse our captured dataset of \recroom user reviews. 
From the sentiment analysis, we report whilst currently overall positive that \textit{Rec Room's} mean review sentiment is decreasing. 
From the word/term frequencies, we report in our negative labelled \recroom reviews the frequent use of words/terms referring to: problematic child users, verbal harassment, and racist language - all issues identified in past works investigating harassment in \textit{Rec Room's} community \cite{maloney-overview, fiani2024pikachu, freeman-disturbing-peace, avatar-user-reviews}. 
Finally, from the BERTopic modelling, we report the identification of 8 themes documenting complaints user have in our negative reviews: \textit{A Toxic Community}, \textit{Problematic Child Users}, \textit{Racism}, \textit{Bullying}, \textit{LGBTQ+ Harassment}, \textit{Banned Users}, \textit{A Faulty Reporting System}, and \textit{Technical Issues} - issues past research has also identified in \textit{Rec Room's} community \cite{blackwell-harassment-long-paper, maloney-child-1, blackwell-harassment-short-paper, freeman-body-avatar-me, mum-user-reviews}.

\subsection{Reflecting on our Analysis Approach}
Reflecting on our analysis, the natural language processing techniques we used provided good high level insights such as monitoring the sentiment of reviews over time and identifying issues raised within our negative reviews. 
Focusing on our BERTopic analysis, the depth of insight obtained from our approach for each of the identified themes does differ from other analysis approaches often used in user experience research, e.g. a manual thematic open-axial-selective coding approach \cite{openaxebook}.
That is, use of our BERTopic approach alone does not provide detailed accounts of the specific issues faced by users in the identified themes/clusters. 
Instead, our approach clusters reviews based on detected similarities between their texts, and provides researchers with keywords summarises the detected similarities in a cluster. 
For example, \textit{``racists, racist users, racism''} is one set of such keywords associated with an identified cluster.  
From these keywords the theme of many clusters can be inferred, in particular if combined with analysis conducted alongside the BERTopic modelling (e.g. sentiment analysis or word/term frequency and importance analysis), and awareness of the subject of the analysis (e.g. for our study awareness of harassment in social VRs). 
Furthermore, if a more detailed understanding of a particular theme/cluster is wanted then the researchers can conduct follow-up analysis with alternative methods (e.g. a manual open-axial-selecting coding \cite{openaxebook}) on that theme/cluster's reviews.

We argue then that our analysis approach provides an effective method of automatically reducing a large dataset of reviews into thematic groupings of reviews based on their similarity to each other. 
This provides researchers with a high level overview of the dataset's contents, and enables more targeted analyses to be conducted if desired by the researchers. 
It also provides a means of transforming a dataset that is too large to feasibly inspect/analysis manually (e.g. a dataset containing 40,000+ reviews) into clusters of reviews which can be manually inspected/analysed (e.g. a cluster of 162 reviews, 205 reviews, 98 reviews, etc).

Beneficial also is the platform agnostic nature of our approach. 
For example, if future work sought to explore a different social VR platform then it is simply a matter of passing a different URL to the data collection script and re-running the analysis scripts. 
The speed and ease of this process is advantageous over alternative methodologies, e.g. it would require significantly more time and resources to re-run an interview of in-situ observation study conducted for one social VR platform on another. 
Furthermore, our analysis also enables new types of cross-platform comparison to be made. 
Real-time sentiment monitors could be established to compare the sentiment of user reviews of many social VRs on a weekly, monthly, or yearly basis, highlighting abnormal shifts due to some change made or rise in harassment in a social VR community. 
For consumers such information could be used to guide their use of these platforms.
Consider the parent trying to ensure their child's safety online - such information could help them to select between social VR platforms or to monitor the safeness of the online spaces their children use \cite{fiani2024pikachu, fiani2022investigating}. 
And for researchers such information could be used to identify abnormal shifts in sentiment with these platforms as they occur and be used to direct investigatory work to understand the cause of such shifts \cite{mum-user-reviews}.

\subsection{Data Collection Beyond User Reviews and Establishing Platform Agnostic Reporting}
We used sentiment analysis, word/term frequencies, and BERTopic modelling for our analysis approach but alternative approaches do exist and parameters used in our approach could be adjusted - which may generate more (or less) accurate results. 
While experimentation with analysis approaches is encouraged in follow up work, future work may also consider how alternative data collection might be developed alongside monitoring mechanisms to identify emergent issues. 
Our approach attempted to piggyback off existing text reviews due to their widespread use use as a reporting mechanism by users \cite{avatar-user-reviews}. 
However, the contents of these reviews is known to vary considerably from high level insights to more descriptive accounts of experienced issues \cite{mum-user-reviews}.
Future work could therefore examine how user reviews on storefronts might be redesigned to better support more structured reviews, e.g. through incident reporting techniques \cite{ismar-stories} to capture more structured accounts of experiences being documented, or by separating different considerations such as the application's design, the experience with any online community, experienced technical issues, etc. 
Future work might also aim to develop dedicated reporting systems to capture user accounts and perspectives, possibly even with video evidence, of experienced issues within social VR platforms. 
Such changes to the source data would aid the development of monitoring tools for social VRs by providing higher quality data to analyse which would produce more accurate insights into these platforms.   
Work could also look to establish a platform agnostic reporting scheme for users of social VRs - where users reports are logged regardless of application or platform in use - as part of efforts to establish independent metaverse/social VR user protection groups \cite{mum-rights, joseph-viewpoint} that are in place to hold developers of social VR platforms accountable for the experience and safety of their users.

%% file: tex_files/related_work.tex
\section{Related Work}

\subsection{Harassment in Social VR Platforms}
Users of social VR platforms, like any online \cite{women-harassment-online-games} or physical \cite{imwut-joseph, chi-joseph} space, are at risk of experiencing harassment from others \cite{maloney-overview, bystander-perdis-joseph, shady-bans}.
While expected given the long history of user harassment in online communities (e.g. as witnessed in online games \cite{mmo-harassment-2004}, social media \cite{pater2016characterizations}, online comments \cite{golbeck2017large}, etc), social VR presents new challenges. 
Because users of social VR often immerse themselves in with identity/avatar on these platforms prior work has demonstrated how such embodiment and immersion can amplify experienced harms such as harassment \cite{freeman-body-avatar-me}. 
Furthermore, the potential types of harassment that social VRs enable both mimics existing platforms (e.g. mobbing, verbal/textual abuse \cite{griefer-taxonomoy}) but also exceeds it, enabling new forms of more intense harassment to occur (e.g. avatar rapes \cite{avatar-rape}, abusive augmentations \cite{jolie-vrst, avatar-user-reviews, maloney-overview}, etc). 

Research is necessary then to investigate how harassment occurs and evolves on social VR platforms within their online communities \cite{mum-user-reviews}. 
To this end, much research has been conducted into harassment on these platforms through user surveys \cite{ieee-white-paper}, expert focus groups \cite{fiani2024pikachu}, interviews \cite{teenagers-interview}, in-situ observations \cite{maloney2020talking}, or a combination of these research methods \cite{maloney-overview}. 
Such works have documented many instances of harassment in social VR with Blackwell et al. reporting experienced harassment can be categorised as: verbal harassment (e.g. insults, slurs, taunting, screaming, etc), physical harassment (e.g. unwanted touching, space violations, etc), or spatial harassment (e.g. presenting unwanted content in a shared space, etc) \cite{blackwell-harassment-short-paper, julie-social-vr-2}. 
Others, meanwhile, have investigated who the target of such harassment are with Freeman et al. identifying  marginalised groups within social VR contexts (e.g. non-male genders,  LGBTQ, and ethnic minorities) are subject to group-based targeting \cite{freeman-disturbing-peace}, a result also found by Blackwell et al. who reported identity cues are used as a targeted form of harassment (e.g. users making sexist remarks to a female avatar) \cite{blackwell-harassment-long-paper}.

Work has also investigated specific user groups within social VR communities.
Schulenberg et al. investigated how women experience and manage harassment during social VR user \cite{women-social-vr} and Maloney et al investigated adult-child interactions within these platforms, focusing on the risks posed to children (e.g. grooming by adults, being exposed to age inappropriate content/conversations, etc) \cite{maloney-child-1, maloney-child-2}. 
Fiani et al have focused on the later in their work \cite{fiani2024pikachu, big-buddy, fiani2022investigating} and investigated the development of safeguarding and protection mechanisms for these platforms, motivated by context aware solutions capable of interpreting a user's experience and reacting accordingly \cite{reality-aware, avi-joseph, imx-joseph}.
However, while prior research presents a clear case for the existence of harassment in social VRs \cite{maloney-overview}, the methods employed to conduct this research are high cost (e.g. in-situ observations, interviews, etc) \cite{mum-user-reviews} and do not allow for the continued monitoring and detection of emergent harms in social VR platforms in platform agnostic manner.

\subsection{Analysis of User Reviews}
One proposed approach of creating platform agnostic automated monitoring systems to detect emergent harms in social VR platforms is the analysis of user reviews posted about the platform \cite{mum-user-reviews, avatar-user-reviews}. 
Such an analysis can not only enable the real time monitoring of emergent threats but also provide longitudinal analyses that explore how user behaviours change over time (e.g. in response to changing user moderation approaches implemented within a social VR platform), and provide insight into the prevalence of different harassment types \cite{mum-user-reviews, avatar-rape}.  

User reviews are a promising source of data for such analyses as they existing in large quantities (e.g. \textit{Rec Room} a popular social VR platform having 40,000+ reviews posted on its \textit{Steam} storefront page), are dated by their nature (allowing for time based explorations to be performed), and are one of the few independent, publicly available insights into the activities that occur on these platforms \cite{user-review-tax}. 
User reviews are also known to include a variety of topics of interest to researchers including: usability \cite{user-review-detection} and accessibility issues \cite{accessibility-reviews}, insights into ethical concerns surrounding an application's use \cite{user-reviews-depression}, emotional content \cite{bertopic-emotions}, or indications of changing player behaviours \cite{topic-modelling-pokemon-go}, etc. 
They are also often used as a form of protest/complaint with users attempting to invoke change through the pressure such posts exert on the platform owners \cite{user-review-tax}. 

In the context of social VRs, O'Hagan et al manually thematically analysed 1000 \textit{Rec Room} user reviews to determine their suitability for large scale analysis \cite{mum-user-reviews}. 
They found that user reviews are used to report user accounts of witnessed and/or experienced harassment in social VRs with varying degrees of detail describing the incidents reported, and conclude user reviews have good potential as a data source for such large scale analyses. 
Building on this work, Dong et al explored avatar experiences reported in social VR user reviews using a dataset of approximately 6000 user reviews and a targeted, natural language processing analysis approach \cite{avatar-user-reviews}. 
They found emergent themes in user reviews discussing avatar customisation, diversity, theft, and ``crashers'' - the latter of which refers to user avatars which \textit{``upload a cache, script, polygon heavy avatar to the VRChat to overload people’s worlds and viewers and crash them, causing massive frustration to players''} \cite{avatar-user-reviews}, and is a problem not reported as occurring in \textit{Rec Room} user reviews \cite{mum-user-reviews}. 
Dong et al's use of a natural language processing analysis approach also provides initial insights into the suitability of such an approach, providing evidence of the effectiveness of topic modelling analysis of user reviews of social VR platforms. 

%% file: tex_files/conclusion.tex
\section{Conclusion}
We conducted an initial exploration of analysing \textit{Rec Room} user reviews using natural language processing techniques.
We captured a dataset of approximately 40,000 \textit{Rec Room} user reviews from the \textit{Steam} storefront dating from June 2016 up to and including February 2024. 
We conducted a sentiment analysis of this dataset, showing user sentiment towards \textit{Rec Room} is decreasing over time, and to automatically label each review in our dataset's sentiment. 
We then conducted follow up analysis on these identified negative reviews. 
First, we investigated word/term frequencies within then, identifying frequent reference to children, verbal harassment, and racism in the reviews. 
We then used BERTopic modelling to thematically group our reviews based on detected similarities between their content, identifying 8 themes concerning emergent issues within \textit{Rec Room's} community.  
We close by reflecting on our analysis approach, highlighting how it might be used effectively by researchers of social VR platforms and identifying areas future works might build on whilst working towards the development of platform agnostic monitoring systems for social VR platforms.